\documentstyle[12pt]{article}
\begin{document}
\title{Conformal blocks in the QCD Pomeron formalism}
\author{
H. Navelet and R. Peschanski\\CEA, Service de Physique
Th\' eorique, CE-Saclay\\ F-91191 Gif-sur-Yvette Cedex, France}
\maketitle
\begin{abstract}
The conformal invariance properties of the QCD Pomeron in the transverse
plane allow us to give an explicit analytical expression for the conformal eigenvectors in the mixed representation in terms of two conformal blocks, each block being the product of an holomorphic times an antiholomorphic function.  
This property is used to give an exact expression for various functions of 
interest, the Pomeron amplitude in both momentum and impact-parameter variables, the QCD dipole multiplicities and dipole-dipole cross-sections in the whole parameter space, and we recover the expression of the four-point gluon Green function given recently by Lipatov.
\end{abstract}
\bigskip
{\bf 1.} Introduction
\bigskip

In his inspiring study \cite{lip} Lev Lipatov has shown that the equation obeyed by the BFKL kernel \cite{BFKL} of the   bare QCD Pomeron is invariant by the (global) conformal group of transformations in the tranverse coordinate space. Using a complete basis of conformal eigenfunctions $E^{n,\nu},$ he is able to express the elastic off-mass-shell gluon-gluon amplitude as an expansion over this basis, when $n,$ the conformal spin, is an  integer and $\nu$ corresponds to a continuous imaginary scaling dimension. In order to investigate the physical properties of this expansion, he is led to consider the eigenfunctions in a mixed representation, $E^{n,\nu}_q,$ which is obtained by a suitable Fourier transform 
of the $E^{n,\nu}.$ Given the nodal importance of these $q-$dependent eigenfunctions for the determination of the solutions of the BFKL equation and of their properties, it appears useful to go a step further. There already exists numerical \cite {salam} and various approximate analytic \cite {lip,muella,samuel} estimates of these quantities.  The aim of our paper
is to derive the exact analytical expressions for the elements of this basis. We perform this derivation by using a powerful method taking advantage of the conformal invariant properties of the theory. 

To be more specific, we find that the eigenfunctions  $E^{n,\nu}_q$ exhibit
 a structure of {\it conformal blocks} which already appeared, in particular, in the computation of correlation functions in 2-dimensional conformal-invariant quantum field theories\cite{correl}.  This structure is a finite linear combination of functions factorized in terms of holomorphic and antiholomorphic parts. The coefficients of this combination are such that they preserve the singlevaluedness of the physical quantities in the complex plane of the coordinates. This structure is then reproduced under different forms in all 
the relevant quantities in the QCD Pomeron calculations. 

In section 2 we give the expression of the $E^{n,\nu}_q$ in terms of their two conformal blocks. Then, in section 3, we derive the analytical expression in conformal blocks for 
$f_\omega (k,k',q)$ which can be interpreted \cite{lip} as the t-channel partial-wave amplitude for gluon-gluon scattering with gluon virtualities $-(k)^2,$ $-(k')^2,$ $-(q-k)^2,$ $-(q-k')^2;$ $t=-q^2$ is the momentum transfer and $\omega$ is the the Mellin-conjugate of the c.o.m. energy squared $s.$ In the next section 4 we give an exact expression of the dipole multiplicity which has been recently introduced and used in an asymptotic approximate form by Al Mueller \cite{muella}. It turns out that this quantity, which coincides with the Pomeron Green function recently calculated by Lipatov \cite{lipa}, obeys the conformal-block structure. In the last section 5, we determine 
the impact-parameter-dependent amplitude, first in the Lipatov original formulation and second in the Mueller approach and exhibit the non-trivial mathematical property which proves their identity in the whole parameter-space.

\bigskip
{\bf 2.} Conformal blocks for eigenfunctions
\bigskip

Using the notations of Ref. \cite{lip}, the eigenfunctions in coordinate space
are defined as:

\begin{equation}
E^{n,\nu}\left(\rho_{10},\rho_{20}\right) = (-1)^n \ \left(\frac {\rho_{10}\rho_{20}}{\rho_{12}}\right)^{\mu-1/2}
 \left(\frac {\bar \rho_{10}\bar \rho_{20}}{\bar \rho_{12}}\right)^{\tilde \mu-1/2} ,
\label{1}
\end{equation}
where $\rho_{ij}\equiv \rho_i-\rho_j$ and $\rho_i \ (i=0,1,2)$ are complex transverse coordinates. The conformal dimensions are defined as :
\begin{equation}
\mu = n/2 -i\nu\ ;\ \tilde \mu =-n/2 -i\nu,
\label{2}
\end{equation}
where $n,$ the conformal spin, is an integer. These are the eigenfunctions of the two Casimir operators of the conformal algebra, namely:
\begin{eqnarray}
\rho_{12}^2 \partial _1 \partial _2 E^{n,\nu}&=&\lambda _{n,\nu} E^{n,\nu} 
\nonumber \\
\bar \rho_{12}^2 \bar \partial _1 \bar \partial _2 E^{n,\nu}&=&\lambda _{n,-\nu} E^{n,\nu}
\nonumber \\
\lambda _{n,\nu} = 1/4 - \mu^2&,&\ \lambda _{n,-\nu} = \bar \lambda _{n,\nu}.
\label{3}
\end{eqnarray}
Lipatov introduces the following mixed representation of the eigenfunctions:
\begin{equation}
E^{n,\nu}_q (\rho) = \frac {2\pi^2}{b_{n,\nu}} \frac {1} {\vert\rho\vert}
\int {dz d\bar z\  e^{\frac {i}{2} (\bar q \rho + q \bar \rho)}\ 
E^{n,\nu}\left(\! z\! +\! \rho/2, z\! -\! \rho/2\right)}\ ,
\label{4}
\end{equation}
where
\begin{equation}
b_{n,\nu}=\frac {2^{4i\nu}\pi^3} {\vert n\vert/2 -i\nu} \
\frac {\Gamma (\vert n\vert/2\! -\!i\nu\!+\!1/2) \Gamma (\vert n\vert/2\! +\!i\nu)}
{\Gamma (\vert n\vert/2 \!+\!i\nu\!+\!1/2) \Gamma (\vert n\vert/2 \!-\!i\nu)}.
\end{equation}

Using the eigenvalue equations (\ref{3}), we obtain the corresponding differential equations obeyed by the $E^{n,\nu}_q.$
\begin{eqnarray}
\left\{\frac {\partial^2}{\partial y^2} + \frac 1y \frac {\partial}{\partial y}
+\left(1-\frac {\mu^2}{y^2}\right)\right\} E^{n,\nu}_q (\rho) = 0,\nonumber \\
\left\{\frac {\partial^2}{\partial \bar y^2} + \frac 1 y \frac {\partial}{\partial \bar y}
+\left(1-\frac {\tilde \mu^2}{\bar y^2}\right)\right\} E^{n,\nu}_q (\rho) = 0,\label{5}
\end{eqnarray}
where $y=\bar q \rho/ 4 .$  Each of these equations admits two linearly independant solutions which are $\bar q ^{\mp \mu} J_{\pm \mu} (y)$ and $ q ^{\mp \tilde \mu}J_{\pm \tilde \mu} (\bar y).$ So, the generic solution for  $E^{n,\nu}_q$ reads
\begin{equation}
 E^{n,\nu}_q (\rho) =  \bar q ^{-\mu} q ^{- \tilde \mu}\ \sum _{\alpha = \pm \mu,\beta = \pm \tilde \mu} C_{\alpha,\beta}\  J_\alpha (y)  J_\beta (\bar y),
\label{6}
\end{equation}
where $C_{\alpha,\beta}$ are constants. Now, the requirement 
 is that  the solution be a monovalued function with respect to the complex variable $\rho.$ This implies that only the combinations $(\mu,\tilde \mu)$ and 
$(-\mu,-\tilde \mu)$ contribute, in order to match the phases of the Bessel functions for each product. This finally yields:
\begin{equation}
 E^{n,\nu}_q (\rho) =  \bar q ^{-\mu} q ^{- \tilde \mu}\ \left[ C_{\mu, \tilde \mu}  J_{\mu} (y)  J_{\tilde \mu}(\bar y) +  C_{-\mu, -\tilde \mu}  J_{-\mu} (y)  J_{-\tilde \mu}(\bar y)\right].
\label{7}
\end{equation}
We determine the coefficients by matching with the known behaviour \cite {lip} when $\vert \rho q\vert \simeq 0.$ We get at once 
\begin{equation}
 \frac {C_{\mu, \tilde \mu}} {  C_{-\mu, -\tilde \mu}} = -1 ;\  C_{-\mu, -\tilde \mu} = 2^{-6i\nu} \Gamma \left(1-i\nu +\vert n \vert/2 \right) \Gamma \left(1-i\nu - \vert n \vert/ 2 \right).
\label{8}
\end{equation} 
This finally yields:
\begin{eqnarray}
 E^{n,\nu}_q (\rho) &=&  \bar q ^{i\nu - n/2} q ^{i\nu + n/2}\ 
 2^{-6i\nu} \Gamma \left(1\!-\!i\nu \!+\!\vert n \vert /2 \right) \Gamma \left(1\!-\!i\nu \!- \!\vert n \vert /2 \right)\times \nonumber \\&\times&
\left[  J_{n/2 -i\nu} (\frac {\bar q \rho} 4)  J_{-n/2-i\nu}(\frac {q \bar \rho} 4) -  J_{-n/2 + i\nu} (\frac {\bar q \rho} 4)  J_{n/2 +i\nu}(\frac {q \bar \rho} 4)\right].
\label{9}
\end{eqnarray}
The form obtained in formula (\ref {9}) is the same as  the one obtained with the conformal block structure of correlation functions in 2-dimensional conformal field theories \cite {correl}. It corresponds to the holomorphic factorization of the integrand in expression (\ref {4}). Note that this integrand is not of the form $f(z) \bar f(\bar z)$ but $f(z) g (\bar
z),$ where the singularity structure of f and g are matched because the conformal spin $n$ is integer.  The Bessel functions appearing in formula 
(\ref {9}) come from the known contour integral representations  \cite {gradstein}
\begin{equation}
H_{\sigma}^{(1,2)} =  \frac{\Gamma (1/2 -\sigma)\ (y/2)^{\sigma}}{i\pi^{3/2}}
\int_{{\cal C}_{(1,2)}}\ e^{izt} \ (z^2-1)^{\sigma-1/2}\ dz,
\label{10}
\end{equation}
where the two Hankel functions $H_{\sigma}^{(1,2)}$ are obtained with the appropriate contours around the singularities, see \cite {gradstein}. The precise combination appearing in (\ref {9})  comes from the analyticity properties of the solution.

\bigskip
{\bf 3.} Computation of the QCD Pomeron amplitudes
\bigskip

Formula (\ref {9}) gives directly the expression of the t-channel Pomeron partial wave amplitude $f_\omega ^q (\rho,\rho')$ in the mixed representation \cite {lip}. let us define  
\begin{equation}
f^{n,\nu}_q (\rho,\rho') = \frac {\vert\rho \rho'\vert} {16} \ \bar 
E^{n,\nu}_q (\rho')\ E^{n,\nu}_q (\rho) \left(\left[\nu^2\! +\!\left(\frac {n\!-\!1} 2\right)^2\right]
\left[\nu^2\! +\!\left(\frac {n\!+\!1} 2\right)^2\right]\right)^{-1},
\label{11}
\end{equation}
where 
\begin{equation}
f_\omega ^q (\rho,\rho')= \sum_{n= -\infty}^{+\infty}\ \int_{\nu = -\infty}^{+\infty}\ d\nu f^{n,\nu}_q (\rho,\rho')\ \frac {1}{\omega -\omega(\nu,n)},
\label{12}
\end{equation}
and \cite {lip}
\begin{equation}
\omega(\nu,n)= \frac {2 \alpha_S N_c} \pi \left(\Psi (1) - Re \left\{\Psi \left(\frac {\vert n\vert + 1 } 2 + i \nu \right)\right\}\right).
\label{13}
\end{equation}
Analogously, one can introduce the corresponding amplitudes in the impact-parameter space, namely $f_\omega (\rho_1,\rho_2,\rho_1',\rho_2')$ and 
$f^{n,\nu} (\rho_1,\rho_2,\rho_1',\rho_2').$ These functions are the Fourier-transforms of the previous ones (\ref {12},\ref {13}), namely:
\begin{eqnarray}
f^{n,\nu} (\rho_1,\rho_2,\rho_1',\rho_2') = \int\  d^2q \ e^{-iq \frac {\rho_{11'} +
\rho_{22'}} 2}\ \frac {\vert\rho \rho'\vert} {16}\ \bar E^{n,\nu}_q (\rho') \ E^{n,\nu}_q (\rho)\nonumber \\ \times \ \left(\left[\nu^2\! +\!\left(\frac {n\!-\!1} 2\right)^2\right]
\left[\nu^2 \!+\!\left(\frac {n\!+\!1} 2\right)^2\right]\right)^{-1}.
\label{14}
\end{eqnarray}
Finally, one defines the amplitudes in the momentum space by \cite {lip}
\begin{eqnarray}
\delta^2 (q-q') f^{n,\nu} (k,k',q) = (2\pi)^{-8} \int  d^2\rho_1 d^2\rho_2 d^2\rho_1' d^2\rho_2'\times \nonumber \\ \times e^{ik\rho_1 +i (k-q) \rho_2 -i k'\rho_1' -i(q'-k')\rho_2'}
f^{n,\nu} (\rho_1,\rho_2,\rho_1',\rho_2'),
\label{15}
\end{eqnarray}
where $k,$ $k',$ $q-k$ and $q'-k'$ are the transverse momenta of the external off-shell gluons (with propagators included).

Let us now derive the analytic solution for $f^{n,\nu} (k,k',q).$
For this sake, we introduce the new variables of integration:
\begin{equation}
\rho = \rho_1\!-\!\rho_2;\ \rho' = \rho_1'\!-\!\rho_2';\ b= \frac {\rho_{11'} \!+\!
\rho_{22'}} 2;\ \sigma = \frac {\rho_1\!+\!\rho_2\!+\!\rho_1'\!+\!\rho_2'} 2.
\label{16}
\end{equation}
The integration over $\sigma$ gives the expected $\delta^2 (q-q'). $ The result reads:
\begin{eqnarray}
f^{n,\nu} (k,k',q) = \pi (2\pi)^{-8} \int  d^2\rho d^2\rho' d^2b \ e^{i(k-q/2)\rho} e^{-i (k'-q/2)\rho'} e^{iqb} \times \ \nonumber \\
\times \ f^{n,\nu} (\rho_1,\rho_2,\rho_1',\rho_2').
\label{17}
\end{eqnarray}

With the definition (\ref {15}), the integration over $b$ gives:
\begin{eqnarray}
f^{n,\nu} (k,k',q) = \pi (2\pi)^{-6} 
 \left(\left[\nu^2\! +\!\left(\frac {n\!-\!1} 2\right)^2\right]
\left[\nu^2 \!+\!\left(\frac {n\!+\!1} 2\right)^2\right]\right)^{-1}\ 
\times
\nonumber \\
\ \times\ \int  d^2\rho \ e^{i(k-q/2)\rho}\  \frac {\vert\rho\vert} 4  \ E^{n,\nu}_q (\rho) \ \int  d^2\rho'\ e^{i(k'-q/2)\rho'}\  \frac {\vert\rho'\vert} 4 \ \bar E^{n,\nu}_q (\rho'),
\label{18}
\end{eqnarray}
which exhibits the factorization between the integrals over $\rho$ and $\rho'.$ In order to compute
the integral say, over $\rho$ one again makes use of the conformal-block structure. Using expression (\ref{9}) for $ E^{n,\nu}_q (\rho),$ one is led to  integrals of the form:
\begin{eqnarray}
B_{\pm}(\bar q,\bar k)= \int d\rho\ e^{\frac i 2 \rho \left(\bar k - \frac {\bar q} 2\right)}\ \rho^{1/2}\ J_{\pm(n/2 -i\nu)} \left(\frac {\rho \bar q} 4\right)\ \ \ \ \ 
\nonumber \\ 
= \ \frac {\left(\frac {\bar q} {-4 i (\bar k - \bar q /2)}\right)^{\pm (n/2 - i\nu)}}{\left( -\frac i 2 (\bar k -\bar q /2)\right)^{3/2}}\ \  \ \frac {\Gamma(\pm (n/2 - i\nu) + 3/2)}{\Gamma(\pm (n/2 - i\nu) + 1)}\ \times \ \ \ \ \ \ 
\nonumber \\
2F_1 \left( \pm (n/2\! -\! i\nu)\! +\! 3/2, \pm  (n/2 \!- \!i\nu) + 2;\pm (n/2\! - \!i\nu)\! +\! 1; \left( \!1\!-\!\frac{2\bar k}{\bar q}\right)^{-2} \right)
.
\label{19}
\end{eqnarray}

The same equation holds for the quantities $\tilde B_{\pm} (q,k)$ by changing $(n,\bar q, \bar k)$ $\rightarrow $ $(-n,q,k)$ in the previous expression. One obtains again a conformal block structure, for which the single-valuedness condition
implies the following diagonal form:
\begin{eqnarray}
I^{n,\nu}(k,q) \equiv \int \ d^2\rho \  e^{i(k-q/2)\rho} \ \frac {\vert\rho\vert} 4 \ E^{n,\nu}_q (\rho)  =  \ \ \ \ \ \nonumber \\
=2^{-6i\nu} q^{i\nu+n/2}\bar q ^{i\nu-n/2}\ \Gamma(1-i\nu +n/2) \ \Gamma(1-i\nu -n/2)\ \times \nonumber \\ 
\ \times\left[B_+(\bar q,\bar k) \tilde B_+(q,k) -
B_-(\bar q,\bar k) \tilde B_-(q,k)\right]\ \ \ \ \ \ \ \ \ \ \ \  
\nonumber \\
= D_{++}\ _2F_1 \left( 3+ 2n -2 i\nu, 1/2+n/2 -i\nu;1 +n - 2i\nu;  \frac {\bar q}{\bar k}  \right)\ \times  \nonumber \\
\times \  _2F_1 \left( 3- 2n -2 i\nu, 1/2-n/2 -i\nu;1 -n - 2i\nu;  \frac { q}{k}  \right)\ 
 \nonumber \\
- D_{--}\ _2F_1 \left( 3- 2n +2 i\nu, 1/2-n/2 +i\nu;1 -n + 2i\nu;  \frac {\bar q}{\bar k}  \right)\ \times \nonumber \\ \times \ _2F_1 \left( 3+ 2n +2 i\nu, 1/2+n/2 +i\nu;1 +n + 2i\nu;  \frac { q}{k}  \right),
\label{20}
\end{eqnarray}
where the coefficients are given by:
\begin{eqnarray}
 D_{++}=i^n \ 2^{3-2i\nu}\left(\frac k {1-\frac q {2k}}\right)^{n/2+i\nu-3/2} \left(\frac {\bar k} {1-\frac {\bar q} {2\bar k}}\right)^{-n/2+i\nu-3/2} \ \times \nonumber \\ \ \times \ \Gamma(n/2+i\nu-3/2)\ \Gamma(-n/2+i\nu-3/2)\ \ \ \ \ \ \ \ \ 
\nonumber \\
\frac  {D_{--}} {D_{++}}\ = (-1)^n\ 2^{-8i\nu} \left(\frac k {1-\frac q {2k}}\right)^{n+2i\nu} \left(\frac {\bar k} {1-\frac {\bar q} {2\bar k}}\right)^{-n+2i\nu} \ \nonumber \\ 
\frac{ \Gamma(\!-\!n/2\!+\!i\nu\!+\!3/2)\ \Gamma(\!-\!n/2\!+\!i\nu\!+\!3/2) \Gamma(\!n/2\!-\!i\nu\!+\!1)\ \Gamma(\!-\!n/2\!-\!i\nu\!+\!1)} { \Gamma(\!-\!n/2\!-\!i\nu\!+\!3/2)\ \Gamma(\!-n/2\!-\!i\nu\!+\!3/2) \Gamma(n/2\!+\!i\nu\!+\!1)\ \Gamma(\!-\!n/2\!+\!i\nu\!+\!1)}
\label{21}
\end{eqnarray}
The final result for the amplitude $f^{n,\nu} (k,k',q)$ is
\begin{eqnarray}
f^{n,\nu} (k,k',q) &=& \pi (2\pi)^{-6}  \left(\left[\nu^2 +\left(\frac {n-1} 2\right)^2\right]
\left[\nu^2 +\left(\frac {n+1} 2\right)^2\right]\right)^{-1} \ \times\nonumber \\
&\times&I^{n,\nu}(k,q)\ \bar I^{n,\nu}(k',q).
\label{22}
\end{eqnarray}

\bigskip
{\bf 4.} Green functions and QCD dipole multiplicities
\bigskip

In the QCD dipole formalism \cite {muellb, muella}, one defines  the multiplicity of dipoles of (2-dimensional) transverse size $\rho'$ originated from an initial dipole of transverse size $\rho$ in an high-energy onium-onium scattering with a transverse momentum $q$. In terms of the $(n,\nu)$ representation, it reads \begin{equation}
N^{n,\nu}_q = 
 \frac {\vert\rho\vert} {\vert\rho'\vert}\  E^{n,\nu}_q (\rho) \ \bar E^{n,\nu}_q (\rho').
\label{23}
\end{equation}
 The expression for the $ E^{n,\nu}_q$ corresponding to formula (\ref{9}) gives an explicit analytic realization of the dipole multiplicity in the whole phase space.

A  Fourier transform leads to  the multiplicity of dipoles at an impact parameter $b.$ In terms of the $(n,\nu)$ representation, one writes \cite {muella}
\begin{equation}
N^{n,\nu} (\rho,\rho',b) = 
\int  \ dq  d\bar q\  e^{i/2(q\bar b + \bar q b)} \ \frac {\vert\rho\vert} {\vert\rho'\vert}\  E^{n,\nu}_q (\rho) \ \bar E^{n,\nu}_q (\rho').
\label{24}
\end{equation}
 Using again the expression for the $ E^{n,\nu}_q$, see (\ref{9}), and the conformal block structure  of the integrals, we get (for the holomorphic part):
\begin{eqnarray}
A_{\pm}(\rho,\rho',b)&=& \int d\bar q\ e^{\frac i 2 {\bar q} b} \ J_{\pm(n/2 -i\nu)} \left(\frac {\rho \bar q} 4\right)\ J_{\pm(n/2 -i\nu)} \left(\frac {\rho' \bar q} 4\right)
\nonumber \\
&=& Q_{\pm(n/2\!-\!i\nu)\!-\!1/2} \left(\frac {b^2\!-\!\left(\frac {\rho} 2 \right)^2\! -\!\left(\frac {\rho'} 2 \right)^2}{2 \left(\frac {\rho} 2 \right) \left(\frac {\rho'} 2 \right)}\right),
\label{25}
\end{eqnarray}
where $ Q$ is the Legendre function of second kind \cite{gradstein}.
Selecting the combinations which satisfy the usual analyticity requirements, one finally gets
\begin{eqnarray}
N^{n,\nu} (\rho,\rho',b) =\frac {16} {\pi^2 \vert\rho'\vert^2} \ \times \ \ \ \ \ \ \ \ \ \ \ \ \ \ \ \ 
\nonumber\\
\ \times \ \Gamma(\!-n/2\!+\!i\nu\!+\!1)\ \Gamma(\!-n/2\!-\!i\nu\!+\!1) \Gamma(\!n/2\!+\!i\nu\!+\!1)\ \Gamma(\!n/2\!-\!i\nu\!+\!1)\ \nonumber\\
 \times \ \left[Q_{n/2\! -\!i\nu \!-\!1/2}(z) Q_{-n/2\! -\!i\nu \!-\!1/2}(\bar z)\ -\ Q_{-n/2\! +\!i\nu \!-\!1/2}(z) Q_{n/2 \!+\!i\nu \!-\!1/2}(\bar z)\right],
\label{26}
\end{eqnarray}
where
\begin{equation} 
z \equiv \left(\frac {b^2\!-\!\left(\frac {\rho} 2 \right)^2\! -\!\left(\frac {\rho'} 2 \right)^2}{2 \left(\frac {\rho} 2 \right) \left(\frac {\rho'} 2 \right)}\right).
\label{27}
\end{equation}

As a matter of fact, up to a normalization factor this multiplicity is nothing but the gluon Green function recently calculated by Lev Lipatov \cite {lipa}. Indeed, using the following change of variables:
\begin{equation}
x \equiv \frac {\rho_{11'} \rho_{22'}} { \rho_{12} \rho_{1'2'}} = \frac {1-z} 2; \ h=n/2\!+\!i\nu\!+\!1/2;\ \tilde h = \!-\!n/2 \!+\!i\nu\!+\!1/2,
\label{28}
\end{equation}
where $x$ is the well-known complex anharmonic ratio, and the known \cite{gradstein} definition of the Legendre functions in terms of $ _2F_1,$ one recovers the Lipatov result. By definition the Green function reads
\begin{eqnarray}
G_{n,\nu} \equiv \int \ d^2\rho_0 \bar E^{n,\nu}\left(\rho_{1'0},\rho_{2'0}\right) E^{n,\nu}\left(\rho_{10},\rho_{20}\right)
\nonumber \\
=\frac 1 {\pi^4} \vert b_{n,\nu}\vert^2 \vert \rho \rho'\vert \int \ d^2q\ e^{-iq \frac {\rho_{11'} +
\rho_{22'}} 2}  \ \bar E^{n,\nu}_q (\rho') E^{n,\nu}_q (\rho),
\label{29}
\end{eqnarray}
where $\vert b_{n,\nu}\vert^2  = \frac {\pi^6} {\nu^2 + n^2/4}.$ One finally obtains the identity valid in the whole phase space:
\begin{equation}
N^{n,\nu} (\rho,\rho',b) \equiv  \frac {\nu^2 + n^2/4} {\pi^2\vert \rho'\vert^2} \ G_{n,\nu}. 
\label{30}
\end{equation}

\bigskip
{\bf 5.} Dipole-dipole elastic amplitude
\bigskip

In the BFKL formalism the function $f^{n,\nu} (\rho_1,\rho_2,\rho_1',\rho_2'),
$ see (\ref {15}), can be interpreted as the dipole-dipole elastic amplitude
where  dipoles of size $\rho_{12}$ and $\rho_{1'2'}$ collide at  an impact parameter distance $b= \frac {\rho_{11'} + \rho_{22'}} 2.$ It is straightforward to get:
\begin{equation}
f^{n,\nu} (\rho_1,\rho_2,\rho_1',\rho_2') \equiv \ \left(\nu^2 + n^2/4\right) G_{n,\nu} = \pi^2 \vert \rho'\vert ^2
N^{n,\nu} (\rho,\rho',b) .
\label{31}
\end{equation}
In the Mueller formalism \cite{muella}, the same amplitude between the two incoming  dipoles is evaluated in a different way. The two dipoles interact through the cascading of dipoles with decreasing c.o.m.rapidity. In the central region the  dipoles obtained after cascading at a given impact-parameter interact through the elementary gluon-gluon exchange at $q=0.$ In terms of the conformal-invariant formalism one may write:
\begin{eqnarray}
f^{n,\nu} (\rho_1,\rho_2,\rho_1',\rho_2') = \sum_{m,m'} \ \int \ \frac {d^2\sigma_{12} d^2\sigma_{1'2'}} {\vert \sigma_{12}\ \sigma_{1'2'}\vert^2}
\ d^2c\ d^2c' \ \delta (b-c-c')
\nonumber \\
\int d\mu\
N^{m,\mu} (\rho_{12},\sigma_{12},c)
\int d\mu'\
N^{m',\mu'} (\rho_{1'2'},\sigma_{1'2'},c')\ 
f^{n,\nu}_q \left(\sigma_{12},\sigma_{1'2'}\right)\vert_{q=o},
\label{32}
\end{eqnarray}
The elementary amplitude at $q=0$ is obtained from formulae (\ref {10},\ref {12}):
\begin{eqnarray}
f^{n,\nu}_q (\sigma_{12},\sigma_{1'2'})\vert_{q=o} \equiv 
\left(\frac {\sigma_{12}} {\sigma_{1'2'}}\right)
^{-n/2+i\nu} 
\left(\frac {\bar \sigma_{12}} {\bar \sigma_{1'2'}}\right)
^{n/2+i\nu} \times \nonumber \\
\left\{\left[\nu^2 +\left(\frac {n-1} 2\right)^2\right]
\left[\nu^2 +\left(\frac {n+1} 2\right)^2\right]\right\}^{-1}.
\label{33}
\end{eqnarray}
Note that this elementary amplitude at $q=0$ is directly related to  the dipole-dipole elementary cross-section, up to a normalization factor. Indeed, once integrating $f^{n=0,\nu}_{q=0}$ over $\nu$, one recovers the formula used in ref. \cite {muella} for the dipole-dipole elementary cross-section at high energy.

Let us  prove that we recover the same amplitude as in formula (\ref {31}).
Inserting in (\ref {32}) the definition of the multiplicities  (\ref {26}), the convolution in impact-parameter yields:
\begin{eqnarray}
f^{n,\nu} (\rho_1,\rho_2,\rho_1',\rho_2') = \sum_{m,m'}\ \int d\mu\ \int  d\mu'\ \nonumber \\ \times\
\int \ dq\ d\bar q\ e^{\frac i 2 (q \bar b + b \bar q)}\  E^{m,\mu}_q (\rho_{12}) \ \bar E^{m',\mu'}_q (\rho_{1'2'})
\nonumber \\ \times\ 
\int\ d^2\sigma_{12}\ \bar E^{m,\mu}_q (\sigma_{12})\ (\sigma_{12})^{-n/2+i\nu-1}\ 
(\bar \sigma_{12})^{n/2+i\nu-1}
\nonumber \\ \times\ 
\int\ d^2\sigma_{1'2'}\  E^{m',\mu'}_q (\sigma_{12})\ (\sigma_{1'2'})^{n/2-i\nu-1}\ 
(\bar \sigma_{1'2'})^{-n/2-i\nu-1}.
\label{34}
\end{eqnarray}
The integrations over $\sigma_{12}$ and $\sigma_{1'2'}$ can be performed using again conformal-block techniques \cite {guida}. One finds conformal block integrals of the form (for, say,  the holomorphic part)
\begin{equation}
\int\ d\sigma_{12}\ (\sigma_{12})^{-n/2+i\nu-1} \ J_{m/2-i\mu}\left(\frac {\bar q\sigma_{12} } 4\right)\ =\ \frac 1 2 \left(\frac {\bar q} 8\right)^{n/2-i\nu}\ \frac {\Gamma\left(\frac {i\nu - i\mu +m-n} 2\right)}
{\Gamma\left(1\!-\!\frac {i\nu \!+\! i\mu \!-\!m\!-\!n} 2\right)}.
\label{35}
\end{equation}
the result for the  2-dimensional integration over $\sigma_{12}$ reads
\begin{eqnarray}
\int\ d^2\sigma_{12}\ \bar E^{m,\mu}_q (\sigma_{12})\ (\sigma_{12})^{-n/2+i\nu-1}\ 
(\bar \sigma_{12})^{n/2+i\nu-1}
\nonumber \\
=2^{-2} \frac {\Gamma\left(\frac {i\nu - i\mu +m-n} 2\right) \Gamma\left(\frac {i\nu - i\mu -m+n} 2\right)} 
{\Gamma\left(1\!-\!\frac {i\nu \!+\! i\mu\! -\!m\!-\!n} 2\right) \Gamma\left(1\!-\!\frac {i\nu\! +\! i\mu \!+\!m\!+\!n} 2\right)}
 \nonumber \\
\times \  \Gamma \left(1\!-\!i\nu \!+\!\vert n \vert /2 \right) \Gamma \left(1\!-\!i\nu \!- \!\vert n \vert /2 \right)
\nonumber \\
\times \ e^{\frac {i\pi} 2 (i\nu - i\mu +m-n)} \sin\left(\frac {\pi} 2 (i\nu - i\mu +m-n)\right),
\label{36}
\end{eqnarray}
where the last phase factor is required by the correlation between holomorphic and anti-holomorphic integrals due to the conformal-block structure \cite {guida}. This phase factor can also be found by direct computation of the 2-dimensional integral through $_3F_2 $ hypergeometric functions. note that expression (\ref {36}) is $q$-independent, which is crucial for the final result.

Performing the $\sum_{m,m'}$ and the $ \int d\mu\ \int d\mu'$ in formula (\ref {32}), we obtain $\delta_{mn} \delta_{m'n} \delta(i\nu-i\mu)
\delta(i\nu-i\mu')$ and  $\delta_{-mn} \delta_{-m'n} \delta(i\nu+i\mu)
\delta(i\nu+i\mu').$ Indeed, they correspond to the  only relevant poles in the complex plane integration when $\nu$ and $\mu, \mu'$ are real. Note that these contributions are, as expected, diagonal in $m, m'$ and 
$\mu,\mu'.$ Inserting these $\delta$-functions in formula (\ref {34}), we recover at once the formula (\ref {31}). Note that this proven equivalence is valid in the whole dipole phase-space and not only assymptotically.

\bigskip
{\bf 6.} Conclusion and outlook
\bigskip

Using the conformal invariant properties of the BFKL Pomeron, we have been able to get exact analytic formulae for various quantities of interest in this formulation. The main result concerns the expression for the eigenvectors of the BFKL equation in the mixed representation as a sum of two conformal blocks implying Bessel functions of complex index and argument. Among the main consequences, we obtain an exact formulation of the off-shell gluon-gluon amplitude in the physical momentum space. We also obtain an exact expression for the QCD dipole multiplicities and scattering amplitudes, valid in the whole phase-space, proving the full equivalence between the BFKL and QCD dipole formulation for these quantities. As a check we recover the recent result on the gluon Green function obtained in a different way by Lipatov \cite {lipa}.

The fact that exact expressions based on conformal invariance have been obtained for the eigenvectors in the mixed representation, is promising for handling the multi-Pomeron interactions \cite{muella,salam,gavin}.

{\bf Acknowledgments}
Discussions
with Riccardo Guida and Samuel Wallon have been appreciated.

\eject


\begin{thebibliography}{99}
\bibitem{lip}
L.N. Lipatov {\it Zh. Eksp. Teor. Fiz.}{\bf 90} (1986) 1536 (Eng. trans.
{\it Sov. Phys. JETP} {\bf 63} (1986) 904.

\bibitem{BFKL}
L.N. Lipatov, {\it Sov. J. Nucl. Phys.} {\bf 23} (1976) 642;
V.S. Fadin, E.A. Kuraev and L.N. Lipatov, {\it Phys. lett.} {\bf B60} (1975)
50;
E.A. Kuraev, L.N. Lipatov and V.S. Fadin, {\it Sov.Phys.JETP} {\bf 44} (1976) 45, {\bf 45} (1977) 199; 
I.I. Balitsky and L.N. Lipatov, {\it Sov.J.Nucl.Phys.} {\bf 28} (1978) 822

\bibitem{salam}
G.~P. Salam, {\it Nucl. Phys.} {\bf B449} (1995) 589; {\bf B461} (1996) 512;


\bibitem{muella}
A.H.Mueller, {\it Nucl. Phys.} {\bf B437} (199) 107.

\bibitem{samuel}
H. Navelet and S. Wallon, to appear.

\bibitem{correl}
A.M. Polyakov, {\it Zh. Eksp. Teor. Fiz.} { Lett.\bf 12} (1970) 538, {\bf 66} (1974) 23;
A.A. Migdal, {\it Phys. lett.} {\bf B44} (1972) 112;
A.A. Belavin, A.M. Polyakov and A.B. Zamolodchikov,  {\it Nucl. Phys.} {\bf B241} (1984) 333; 
Vl.S. Dotsenko and V.A. Fateev, 
{\it Nucl. Phys.} {\bf B240} FS{\bf12} (1984) 312; 

For a collection of  reprints on conformal invariance, see
{\it Conformal invariance and applications to statistical mechanics}, C. Itzykson, H. Saleur, J.-B. Zuber eds., World Scientific, 1988.
\bibitem{lipa}
L.N. Lipatov {\it Small-x physics in perturbative QCD}, hep-ph/9610276.

\bibitem{gradstein}
I.S. Gradshteyn and I.M. Ryzhik {\it Table of integrals and products}
(Academic Press, Inc., A. Jeffrey ed., 1994)

\bibitem{muellb}
A.H.Mueller, {\it Nucl. Phys.} {\bf B415} (1994) 373; 
A.H.Mueller and B.Patel, {\it Nucl. Phys.} {\bf B425} (1994) 471.

\bibitem{guida}
R. Guida, private communication; R. Guida and N. Magnoli {\it  Tricritical Ising model near criticality}, to appear; S.D. Mathur, {\it Nucl. Phys.} {\bf B369} (1992) 433.

\bibitem{gavin}
A.~H. Mueller and
 G.~P. Salam, {\it Nucl. Phys.} {\bf B475} (1996) 293;
A.~Bialas and  R.~Peschanski, {\it Phys. lett.} {\bf B378} (1996) 302;  {\bf B387} (1996) 405.


\end{thebibliography}
\end{document}